\documentclass[pre,aps,twocolumn,superscriptaddress,longbibliography]{revtex4-1}

\usepackage[dvips]{graphicx}
\usepackage{amssymb,amsfonts,amsmath,bm,bbm}
\usepackage{color}
\usepackage{ulem}
\usepackage[hidelinks]{hyperref}

\begin{document}

\title{
	{The geometric phase and the dry friction of sleeping tops on inclined planes}}

\author{Sven Barthmann}
\affiliation{Experimentalphysik X, Physikalisches Institut, 
  Universit{\"a}t Bayreuth, D-95440 Bayreuth, Germany}

%
%
\author{Thomas M. Fischer}
\email{Thomas.Fischer@uni-bayreuth.de}
\affiliation{Experimentalphysik X, Physikalisches Institut, 
  Universit{\"a}t Bayreuth, D-95440 Bayreuth, Germany}

\begin{abstract}{
We report on the motion of a spinning sleeping top on an inclined plane. Below a critical inclination angle the sleeping tops are force free. The trajectory of a sleeping top on weakly inclined planes in the adiabatic limit is invariant of the angular frequency of the top and thus invariant under a rescaling of the time, however not invariant under time reversal. The stationary trajectory of the sleeping top is characterized by its angle to the in plane horizontal direction. At larger inclinations of the plane the stationary motion of the top becomes unstable and the top accelerates downhill. The behavior points towards a complex law of dry friction of the contact point between the top tip and the material of the inclined plane that depends on a slip parameter. We propose a phenomenological law of dry friction that can explain the relaxation of the top into the sleeping position, the geometric behavior of the top trajectories, and the instability of the stationary motion at larger inclination angles.  
	}
\end{abstract}

\date{\today}

\maketitle

\section{Introduction}

A geometric phase is a phase difference acquired over each period of a cyclic adiabatic process. The geometric phase is a powerful concept in both, quantum\cite{Berry} and classical\cite{Hannay} physics. Geometric phases are connected to the anholomony of non integrable Hamiltonian systems,
geometric phases however also appear under dissipative conditions\cite{Haken}. The geometric phase is accumulated under adiabatic, i.e. slow enough conditions independent of the speed, while the system follows a path in parameter space. The geometric phase also named \textit{Berry}-phase in quantum sytems and \textit{Hannay}-angle in classical systems is invariant under rescaling of the time schedule with which one passes the path in parameter space. It also carries along a gauge freedom of choice of reference points and the concept has been used in high energy particle physics \cite{HighEPHys}, in solid state physics of topological materials \cite{SSP-TI,Hasan,TI}, in the explanation of the rotation of the Foucault-pendulum\cite{Foucault}, in the explanation of the propulsion of active swimmers in low Reynolds number fluids\cite{Wilczek1,Wilczek2}, the propulsion of light in twisted fibers\cite{fiber,Rechtsman}, the propulsion of acoustic \cite{Mao} or stochastic\cite{Murugan} waves, the motion of edge waves in coupled gyroscope lattices\cite{Nash}, the rolling of nucleons\cite{Harland2018} and in the control of the transport of macroscopic\cite{Rossi} and colloidal\cite{tp1,tp2,tp3,Mahla} particles above magnetic lattices. The independence of the geometric phase of the speed, makes the Foucault pendulum rotation independent of the length of the pendulum as well as independent of the value of the gravitational acceleration, similarly the propulsion of an active shape changing swimmer is independent of the viscosity of the embedding fluid. 

Dry friction is a force counter acting relative lateral motion of two solid surfaces in contact. Two non-moving surfaces resist a relative motion with static friction. Static friction is stronger than  dynamic friction of surfaces that are already in relative motion. Coulomb's law of friction states that dynamic friction is independent of the sliding velocity.
Dry friction is a phenomenological description of very complex underlying physical processes occurring at both surfaces. It is therefore not surprising of finding situations where those phenomenological equations fail\cite{Popov}. 

Here we report on experiments on the motion of sleeping tops on inclined planes\cite{STIP1,STIP2,STIP3,STIP4} that exhibits a \textit{Hannay}-like angle of its trajectory, that for some materials of the top and the inclined plane is independent of the spinning rate but not independent of the sign of the spin, and not independent of the material used. It can be explained using a phenomenological generalized expression for the dry friction force of the top with the inclined plane that we postulate to be a homogeneous Galilei invariant function of the velocities at the contact point.        
The velocity independent friction law of Coulomb is clearly violated and we must replace it by a dry dynamic friction force law that depends on the sliding velocity. 

\section{Experiments}
\begin{figure}[t]
	\includegraphics[width=\columnwidth]{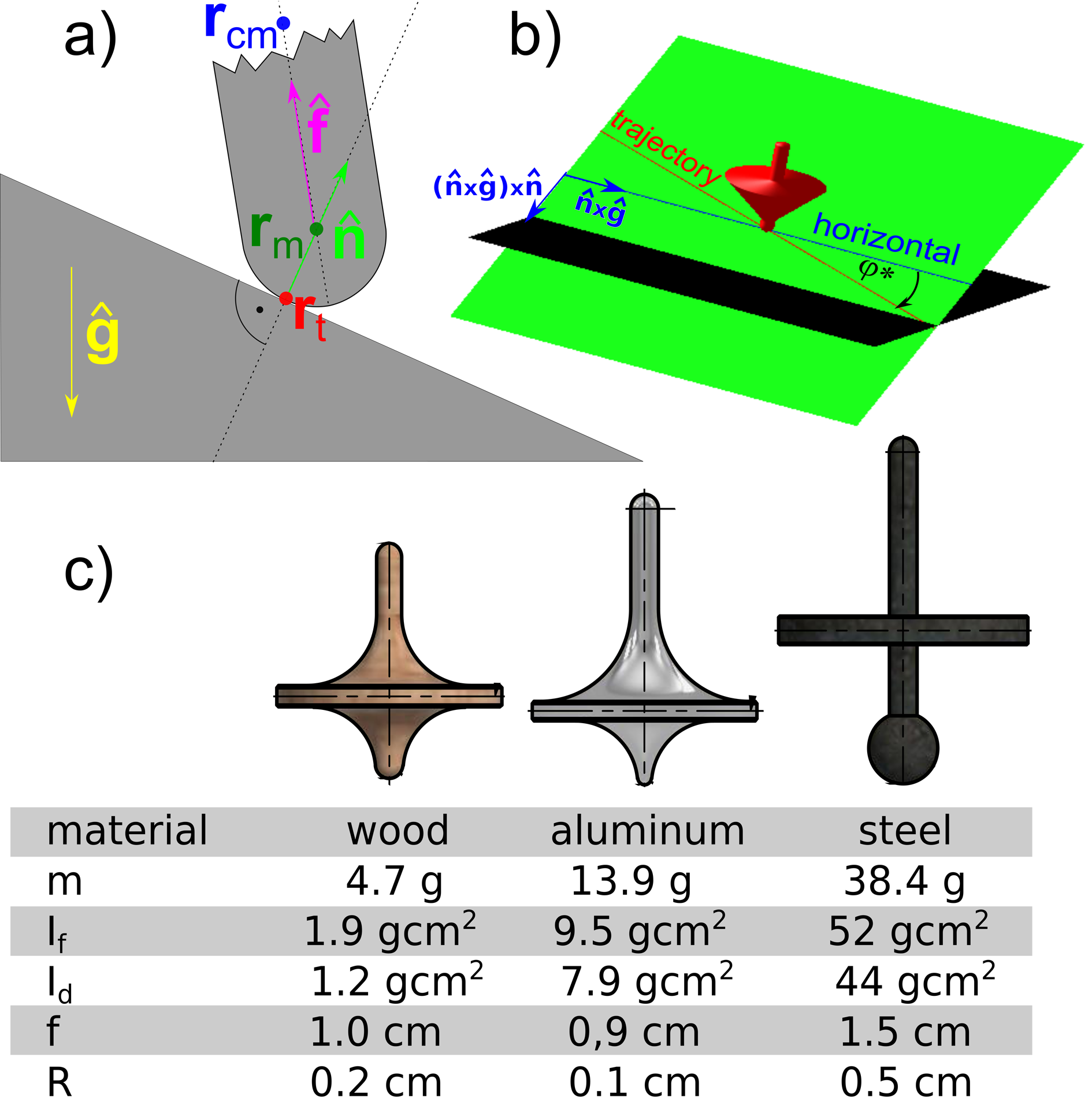}
	\caption{\textbf{a)} Scheme of a spinning top on an inclined plane with touching point $\mathbf r_t$, center of curvature $\mathbf r_m$ and center of mass $\mathbf r_{cm}$, unit vector along gravity $\hat{\mathbf g}$, normal vector $\hat{\mathbf n}$, and figure axis vector $\hat{\mathbf f}$. \textbf{b)} The  trajectory of a force free sleeping top in the stationary state is at an angle $\varphi^*$ to the horizontal of the inclined plane. \textbf{c)} The characteristics of the three tops used in the experiment.\label{figscheme}}
\end{figure} 

We performed experiments with the three different tops each having a roughly spherical base of radius $R$ (see Fig. \ref{figscheme}\textbf{a)}). The midpoint $\mathbf r_m$ of the spherical base is sitting on the position 
$\mathbf r_m=\mathbf r_t+R{\hat{\mathbf n}}$ with ${\hat{\mathbf n}}$ the outward normal to the inclined plane and $\mathbf r_t$ the point where the base of the top touches the inclined plane. The midpoint $\mathbf r_m$ is part of the figure axis of the axisymmetric top and thus the center of mass $\mathbf r_{cm}$ of the top is located at
$\mathbf r_{cm}=\mathbf r_{t} +R{\hat{\mathbf n}} + f{\hat{\mathbf f}}$ with  ${\hat{\mathbf f}}$ the unit vector along the figure axis and $f$ the distance of the center of mass from $\mathbf r_m$. The moment of inertia $\mathbf I=I_d (\mathbbm {1}-{\hat{\mathbf f}}{\hat{\mathbf f}})+ I_f {\hat{\mathbf f}}{\hat{\mathbf f}}$ has the largest eigenvalue $I_f$ along the figure axis and is degenerate $I_d$ in the directions perpendicular to ${\hat{\mathbf f}}$. The properties of each of the three tops are listed in Fig. \ref{figscheme}\textbf{c)}. We  placed the tops on an inclined planes made of either steel or PMMA.  The inclination angle $\alpha$ of the plane is given by $\alpha=\arccos(-{\hat{\mathbf n}}\cdot\hat{ \mathbf g})$, where $\hat{\mathbf g}$ denotes the unit vector in direction of gravity.

\begin{figure}[t]
	\includegraphics[width=\columnwidth]{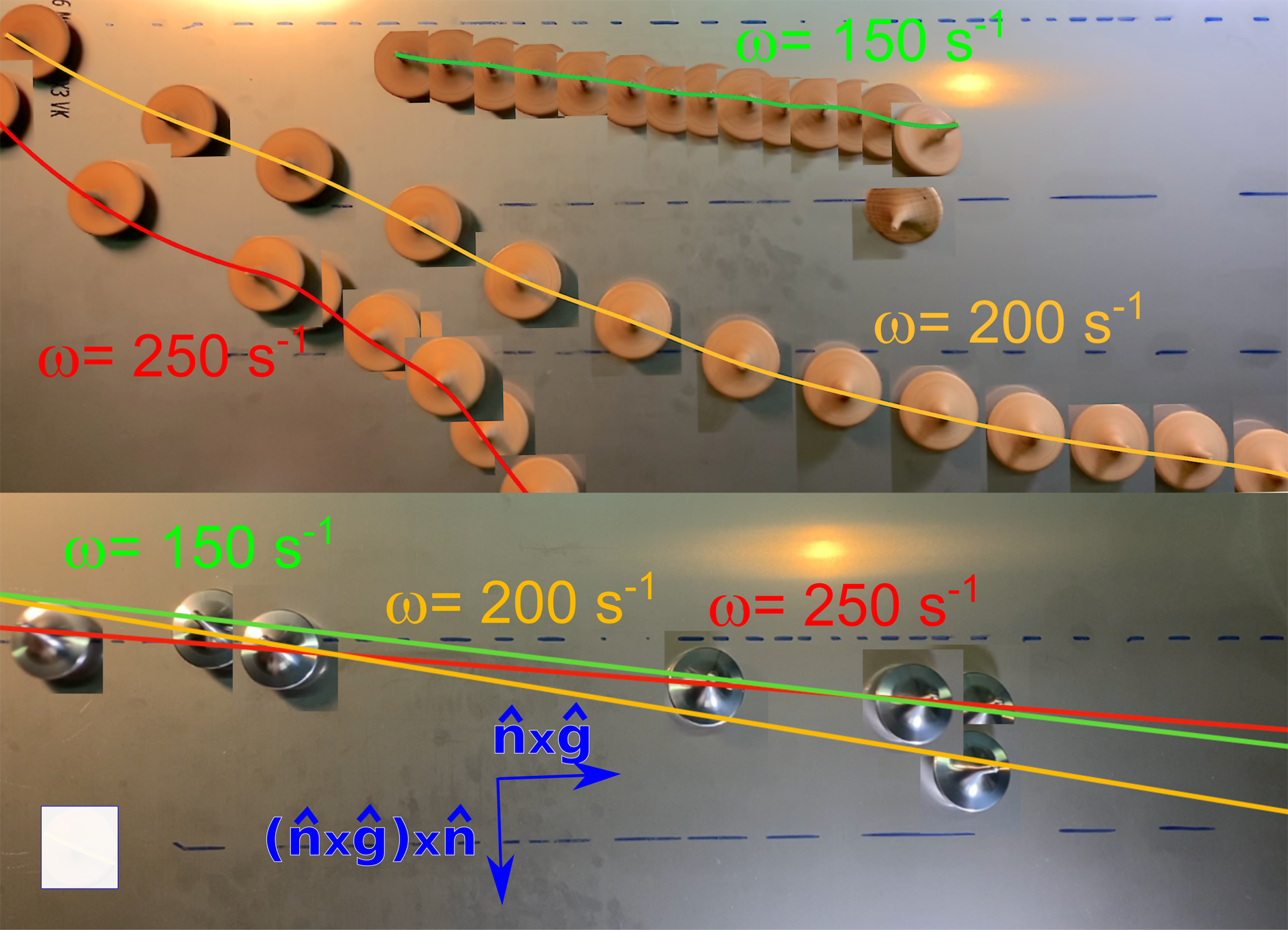}
	\caption{Trajectories of sleeping  tops made of wood (top) and aluminum (bottom) on an inclined $\alpha=0.09$ steel plane. In the adiabatic limit for low angular frequencies the slope of the trajectory of metal tops becomes independent of the angular frequency. The scale box is $32\times 32$ $mm^2$.}  \label{Figadiabatic}
\end{figure}

Fig. \ref{Figadiabatic} shows trajectories with snapshots of the tops on a steel inclined plane with inclination $\alpha=0.09$. Contrary to non spinning objects that exhibit force free and non accelerated steady motion only for one specific inclination angle of the plane ($\tan\alpha=\mu_g$), all tops are moving along force free trajectories of their center of mass at constant speed for all angles below a critical inclination angle $\alpha<\alpha_c=\arctan {\mu_c}$ once they are asleep, i.e. once their figure axis $\hat{\mathbf f}$ and momentary angular frequency $\bm \omega$ are aligned parallel to each other. They reach the sleeping position via a cascade of relaxations first nutating and later precessing before they reach their quiet life\cite{quietlife} behavior. The wooden top shows trajectories that are nonadiabatic, i.e. the slope $\tan \varphi^*$  of the force free trajectory measured with respect to horizontal direction $\hat{\mathbf n}\times {\hat{\mathbf g}}$ of the inclined plane (see Fig. \ref{figscheme}\textbf{b)}) depends on the angular frequency of the top $d\varphi^*_\textrm{wood}/d\omega^*\ne 0$. Only in the adiabatic limit, i.e. for low angular frequencies that not yet render the figure axis instable does the slope of the wooden top become independent of the angular frequency. 
The metal tops are in the adiabatic limit  for all three angular frequencies, with an angular frequency independent slope $d\varphi^*_\textrm{metal}/d\omega^*\approx 0$ of the trajectory.

\begin{figure}[t]
	\includegraphics[width=\columnwidth]{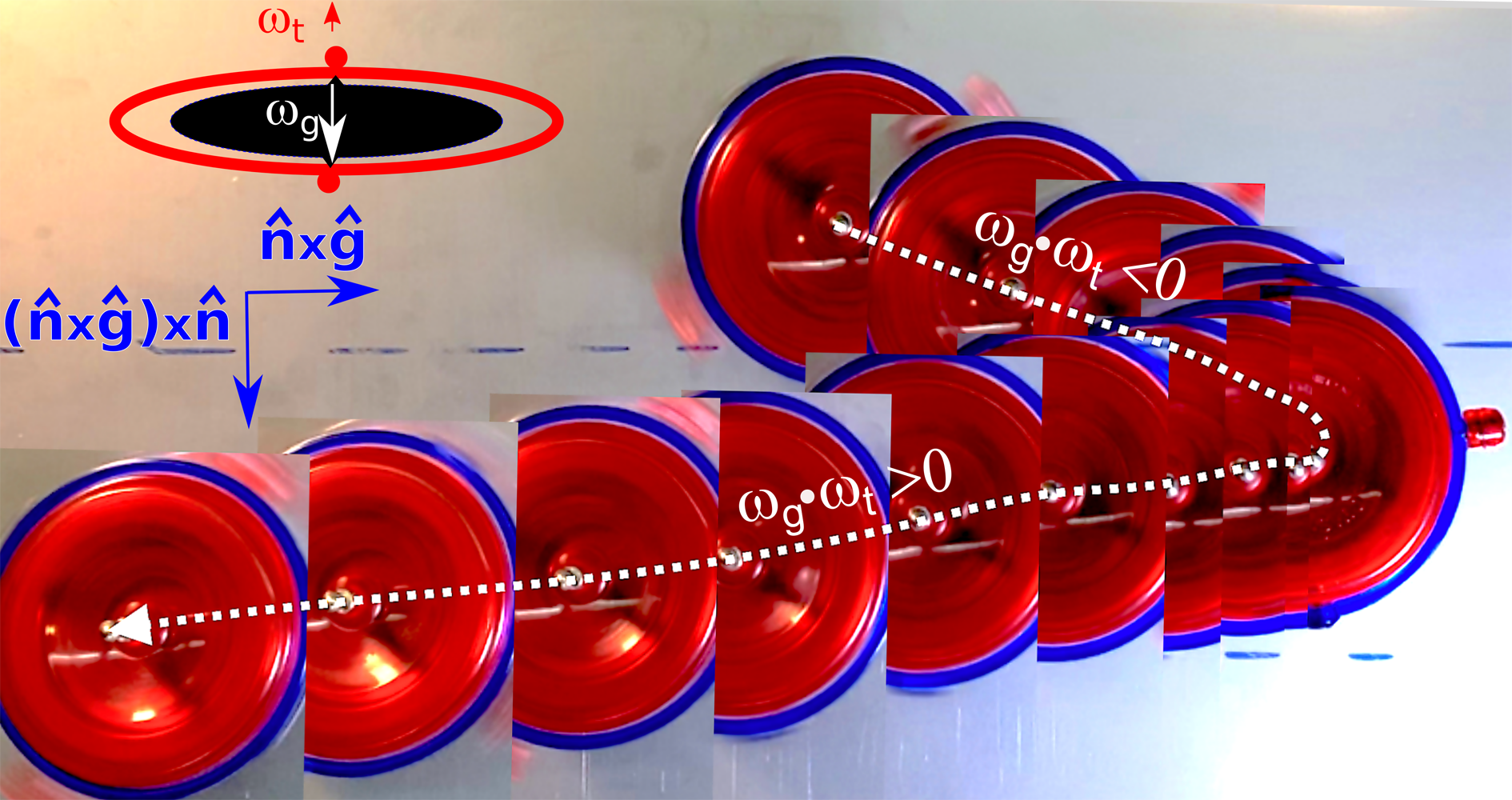}
	\caption{Trajectory of a top (red in the scheme) incorporating a counterrotating gyroscope (black in the scheme) having a steel tip and precessing on an inclined $\alpha= 0.07$ steel plate. The trajectory flips the direction from $\varphi^*_i\to\varphi^*_f=\pi-\varphi^*_i$ while the rotation frequency is slowly reversed by the loaded gyroscope. The behavior is in accordance with equation \ref{flip} of the phenomenological model. }  \label{gyro}
\end{figure}

The dependence of the slope of a top on the angular frequencies can be shown most prominently when using a top rotating at angular frequency $\omega_t$ loaded with a gyroscope rotating in opposite direction at a much larger frequency $\omega_g$  ($0<-\bm{\omega}_t\cdot\bm{\omega_g}/\omega_g^2\ll 1$)  that keeps the figure axis of the top stable even at zero angular frequency of the top $\omega_t=0$. In this configuration the coupling between the top and the loaded gyroscopes lets the angular frequency slowly pass through zero and reverse direction. The resulting trajectory of the gyroscope loaded top is shown in Fig. \ref{gyro}. It does not reverse direction since only the horizontal motion flips while the downhill motion remains the same proving that the problem is not time reversal invariant a property it shares with the behavior of a tippy top\cite{turnover}.

\begin{figure}[h]
	\includegraphics[width=\columnwidth]{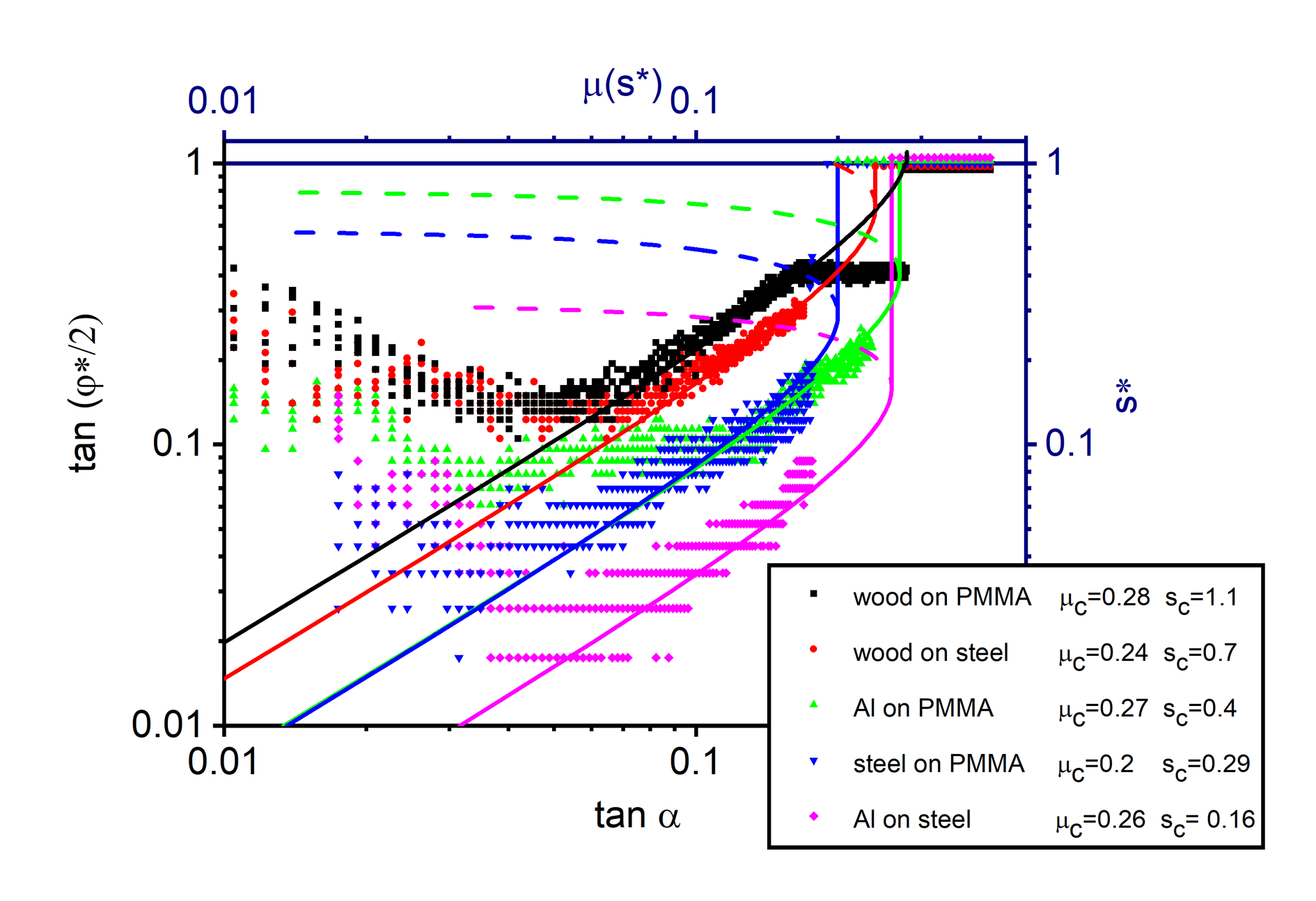}
	\caption{Measurement of $\tan(\varphi^*/2)$ as a function of $\tan \alpha$ for different tops on different planes. The initial angular frequency of the tops were $\omega\approx 200 s^{-1}$. For the metal tops the angular frequencies  fell in a range where the trajectory angle does not depend on the angular frequency $\partial \varphi^*/\partial \omega \approx 0$. The experimental tops exhibit a phase transition with stationary behavior for $\tan\alpha<\mu_c$ and down hill acceleration for $\tan\alpha>\mu_c$. The solid lines are fits to the stable branches of equation \ref{empiricalnonsense} including the dynamic instability near $\tan\alpha=\mu_c$ with parameters indicated in the legend. The dashed lines are the instable branches of equation \ref{empiricalnonsense}. We believe the deviation of the experimental data at very small angles from our model to be due to effects of surface roughness.} \label{Figslopedata}
\end{figure}

In Fig. \ref{Figslopedata} we plot $\tan\frac{\varphi^*}{2}$ as a function of $\tan \alpha$. The experimental data shows stationary geometric behavior with a frequency independent value of $\tan\frac{\varphi^*}{2}$ for $\tan\alpha<\mu_c$. The stationary behavior of the top becomes unstable at $\tan\alpha_c=\mu_c$ and $\tan\frac{\varphi^*}{2}=s_c$. For $\tan\alpha>\mu_c$ there is no stationary behavior and the top rather accelerates downhill along $({\hat{\mathbf n}}\times{\hat{\mathbf g}})\times \hat{\mathbf n}$  with an asymptotic value of $\tan\frac{\varphi^*}{2}=1$. 

%
%

If we have a stable stationary solution that can be described by our model explained in section \ref{sectionmodel} then the tangent of the inclination angle must be equal to the dynamic friction coefficient $\mu_{dyn}(s^*)$ of the model, i.e. $\tan\alpha=\mu_{dyn}(s^*)$. The model friction coefficient depends on the stationary slip parameter $s^*$ which if the model is right must relate to the slope of bisector between the horizontal and the trajectory $\tan(\varphi^*/2)=s^*$. We find a regime where both
$ds^*/d\mu>0$ and $d\varphi^*/d\alpha>0$ where we can fit the data nicely to the model using the empirical equation:
\begin{equation}
\mu_{dyn}(s)=\mu_c \frac{s(2s_c-s)}{s_c^2}\label{empiricalnonsense}
\end{equation}
for the dependence of the friction coefficient on the slip parameter
that predicts $\mu=0$ for zero slip $s^*=0$ and maximal $\mu=\mu_c$ for slip $s=s_c$. 
 The experimental data agrees with the model also in the regime $\mu>\mu_c$ since there $ds^*/d\mu<0$ and the stationary solution is predicted to be unstable and thus the experimental top and the model top accelerates downhill $\varphi^*=\pi/2$ with maximal slip $s=1$. At very low inclination angles $\tan\alpha\ll 1$ the experimental data exhibits negative slope  $d\varphi^*/d\alpha<0$ which is incompatible with our model because it predicts such states to be unstable. 
It is however clear that rolling without slip becomes impossible when the lever arm $R(\hat{\bm{f}}\times\hat{\bm{n}})$ of the top for very low inclination angles of the plane becomes so small that surface roughness prevents the material velocity of the top at the contact point to adapt to the zero material velocity of the inclined plane. We have not incorporated effects of surface roughness into our model. We thus believe this deviations of the experimental data from our simple model to be caused by surface roughness. Surprisingly the behavior of the metal tops nevertheless is geometric e.g. the experimentally measured angle $\varphi^*$ remains independent of the angular frequency $\omega$ of the top. The behaviorat low inclination angles, however, becomes shape dependent and tops with different tip radius $R$ deviate from each other (see the the data of the aluminum and steel top on PMMA). The trajectory of the wooden top is neither angular frequency independent nor does it follow the empirical equation \ref{empiricalnonsense} up to the critical inclination.

\section{Phemomenological model}\label{sectionmodel}

\begin{figure*}
	\includegraphics[width=2\columnwidth]{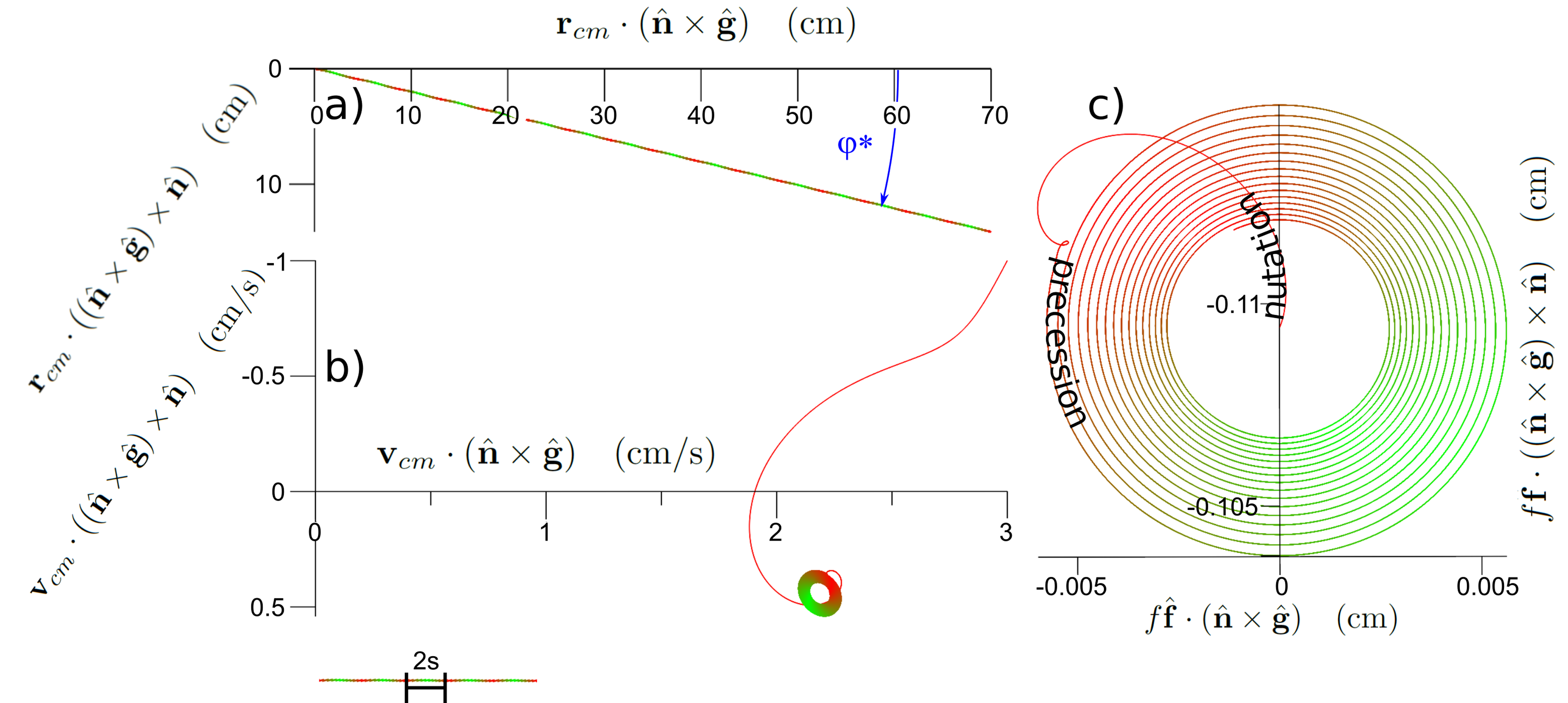}
	\caption{Simulated \textbf{a)} center of mass trajectory $\mathbf r_\textrm{cm}(t)$,
		\textbf{b)} center of mass velocity $\mathbf v_\textrm{cm}(t)$, and \textbf{c)} figure axis $f\hat{\mathbf {f}}(t)$ for the aluminum top of table 1 on an inclined PMMA plane of inclination $\alpha=0.1$ using a linearized friction model $\mu(s)=\mu(s^*)+d\mu/ds \Delta s$ with $d\mu/ds=1$. All trajectories are colored with one second colored in red and the next second in green.  The angular frequency independent \textit{Hannay} type angle $\varphi^*$ is indicated in the trajectory in \textbf{a)}.}  \label{figsim}
\end{figure*}

In order to get a force free behavior of the top for not only one marginal inclination angle, we have to postulate a constitutive equation for the dry friction differing from the standard Coulomb friction law.
We propose a constitutive phenomenological equation of the friction force on a body $1$ sliding and rolling on a support $2$ that reads
\begin{equation}\label{friction}
\mathbf F_\textrm{fr}=-\mu_{dyn}(\hat{\mathbf n}\cdot \mathbf F_{21})\frac{\mathbf v_t^1-\mathbf v_t^2}{|\mathbf v_t^1-\mathbf v_t^2|},
\end{equation}

where $\mu_{dyn}$ denotes the friction coefficient. The expression  $\hat{\mathbf n}\cdot \mathbf F_{21}$ denotes the normal force of the support $2$ onto the body $1$, $\hat{\mathbf n}$ is the outward normal to the support $2$. There are three velocities defining the local dynamics near the contact point:  The first two velocities $\mathbf v_t^i,\, i=1,2$ are the material velocities of the two bodies $i=1,2$ at the contact point. The third velocity is the velocity of the contact point
$d\mathbf r_t/dt$. The difference between the contact point velocity and the material velocity at the contact point  $d\mathbf r_t/dt- \mathbf v_t^i$ is the relative velocity of contact point renewal, i.e. the velocity with which new material points of the surface of body $i$ are explored.  $|d\mathbf r_t/dt-\mathbf v_t^1|+|d\mathbf r_t/dt-\mathbf v_t^2|$ is the speed with which new contacts between the surfaces of both bodies are explored. The key to achieve geometric behavior in the adiabatic limit is to approximate the dynamic friction coefficient to be a homogeneous function of the three velocities: 
\begin{equation}
\mu_{dyn}(\lambda \mathbf v_t^1,\lambda \mathbf v_t^2,\lambda d\mathbf r_t/dt)=\lambda^0 \mu_{dyn}(\mathbf v_t^1,\mathbf v_t^2,d\mathbf r_t/dt)
\end{equation} 

This law is probably violated by the wooden top but should be a good approximation for the metal tops.
If we also assume the friction to be surface isotropic and Galilei invariant, we can write the friction as 
\begin{equation}\label{sliding} 
\mu_{dyn}=\mu_{dyn}(s),\end{equation}
where the Galilei invariant scalar
\begin{equation}
s=\frac{|\mathbf v_t^1-\mathbf v_t^2|}
{|d\mathbf r_t/dt-\mathbf v_t^1|+|d\mathbf r_t/dt-\mathbf v_t^2|},
\end{equation}
describes the ratio of sliding speed with respect to the total speed of contact point renewal.  
If $s=0$ the relative material velocities of both bodies at the contact point coincide such that both bodies roll on each other. For $\mathbf v_t^1=-\mathbf v_t^2$ and $d\mathbf r_t/dt=0$ we have a value of $s=1$ and both bodies slide on a resting contact point into opposite directions. 
Hence our model generalizes the Coulomb law of friction for sliding and rolling to a mixed situation, where both bodies slide and roll at the same time.
For a sliding and rolling spherical surfaced body $1$ of isotropic local curvature $R^{-1}$ on a flat resting  support $2$ ($\bm v_2=\bm \omega_2=\bm 0$)
the velocity of the contact point is $d\mathbf r_t/dt = \mathbf v_t^1+R(\bm{\omega}^1\times \hat{\mathbf n})$, where $\bm \omega^1$ is the angular frequency. The parameter $s$ simplifies to $s={v_t^1}/(R|\bm{\omega}^1\times \hat{\mathbf n}|+|\mathbf v_t^1+R(\bm{\omega}^1\times \hat{\mathbf n})|)$ and we anticipate the friction constant of ideally smooth surfaces to be a function of the slip parameter $\mu_{dyn}=\mu_{dyn}(s)$.   Let us see what this model of friction predicts for a top spinning on an inclined plane.

%
%

  
Since the top is a solid body the velocities of the touching point of the top and the center of mass are related by $\mathbf v_t=\mathbf v_{cm}-\bm\omega\times(R{\hat{\mathbf n}}+f{\hat{\mathbf f}})$, where $\bm \omega=\mathbf I^{-1}\cdot \mathbf L$ is the angular velocity of the top, $\mathbf I$ the moment of inertia and $\mathbf L$ the angular momentum. The gravitational weight $\mathbf F_{cm}=m\mathbf g$ acts on the center of mass of the top of mass $m$. The vector $\mathbf g=g\hat{\mathbf g}$ is the gravitational acceleration with $g$ its magnitude and $\hat{\mathbf g}$ the unit vector in direction of $\mathbf g$. Contrary to the standard law of dry friction we anticipate a friction force $\mathbf F_\textrm{fr}=-\mu_{dyn}(s)(-m\hat{\mathbf n}\cdot \mathbf g)\frac{\mathbf v_t}{v_t}$ according to equation \ref*{friction}. The total force $\mathbf F_t=\mathbf F_\textrm{21}+ \mathbf F_\textrm{fr}$ acting on the touching point therefore is the sum of the normal counterforce of the inclined plane $\mathbf F_{21}=-m{\hat{\mathbf n}}{\hat{\mathbf n}}\cdot \mathbf g$ preventing the top from accelerating into the plane and the unconventional friction force.
We measure the torque $\bm \tau=(\mathbf r_t-\mathbf r_{cm})\times \mathbf F_t$ with respect of the center of mass position.  The equation of motion for the orientation of the top reads $\bm\tau=\frac{d}{dt}\mathbf L$, where $\frac{d}{dt}{\hat{\mathbf f}}=\bm\omega\times {\hat{\mathbf f}}$ 
and the equation of motion of the center of mass position is given by
$m\frac{d}{dt}\mathbf v_{cm}=\mathbf F_{cm}+\mathbf F_t$.

It is straightforward to compute the stationary state for which all time derivatives, forces, and torques vanish. We mark the quantities of the stationary state with a star and they are given by
\begin{eqnarray}
\frac{\mathbf v_t^*}{|\omega^*R\hat f_g^*|}=\tan \frac{\varphi^*}{2}	({\hat{\mathbf n}}\times{\hat{\mathbf g}})\times \hat{\mathbf n}\\
{ {\hat{\mathbf f}}}^*=-\frac{R}{f}{\hat{\mathbf n}} +\hat f_g^*\hat{\mathbf g}\\
\bm\omega^*=\omega^* {\hat{\mathbf f}}^*\\
\frac{\mathbf v_{cm}^*}{|\omega^*R\hat f_g^*|}=\frac{\mathbf v_t^*}{|\omega^*R\hat f_g^*|} +\textrm{sign}(\omega^*)(\hat{\mathbf n}\times {\hat{\mathbf g}}),
\end{eqnarray}

where 

\begin{eqnarray}
\hat f_g^*=\frac{R({\hat{\mathbf n}}\cdot\hat {\mathbf g})}{f}-\sqrt{1-\left(\frac{R({\hat{\mathbf n}}\times \hat{\mathbf g})}{f}\right)^2}
\end{eqnarray}

is the stationary component of $\hat {\mathbf f}^*$ along $\hat{\mathbf g}$
and the stationary angular frequency has an arbitrary magnitude $\omega^*$.
The model predicts:
\begin{itemize}
	\setlength{\itemsep}{.05cm}
	\item a	down hill stationary sliding velocity of the spinning top for inclination angles where a non spinning top sticks to the support.
	\item a horizontal stationary velocity independent of the material of the top and independent of the material of the inclined plane 
	\item a stationary orientation such that the center of mass resides vertically above the touching point
	\item a stationary trajectory of the top that depends only on the friction coefficient, the inclination angle, and the sign of the angular frequency.  
\end{itemize} 

%

The stationary state is stable if $d\mu_{dyn}/ds>0$ and unstable otherwise as can be checked by numerically solving the equations of motion for various sets of parameters or more generally via a linear stability analysis.
Fig. \ref{figsim} shows a simulation of a top not yet at the stationary velocity but slowly approaching it. The spinning top initially nutates, accelerates and then reaches a state where the nutations have died away. In the second stage the top precesses and the velocity spirals around an average velocity of the top. In the final state the top is asleep with the figure axis aligned with the momentary angular frequency and the top rolling along the inclined plane horizontal direction and the same time sliding into the down hill direction, i.e. the travel direction  is with an angle 
\begin{equation}\mu_{dyn}(s^*=\tan\frac{\varphi^*}{2})=-\textrm{sign}(\omega^*)\tan(\alpha)\label{flip}\end{equation} 
to the horizontal direction and the top is moving at constant speed along a path, that is completely independent of the geometric properties and the magnitude of the angular velocity of the top. The top travels more and more downhill the closer the inclination angle is to the critical inclination angle where the velocity of the top diverges. There is no stationary state beyond the critical inclination angle.  

For the metal tops the model seems to predict most but not all characteristics observed in the experiments. The wooden tops behavior is far more complex and out of the scope of the present geometric and thus adiabatic dissipative description.

\section{Conclusions}
Our measurements show a force free motion of sleeping tops on inclined planes at constant velocity for inclination angles below a critical angle. In the adiabatic limit a sleeping spinning metal top moves on a weakly inclined plane along a trajectory that is invariant under rescaling of time, but not invariant under time reversal. The trajectory in contrast to the trajectories of other dissipative systems such as shape changing swimmers in low Reynolds number fluids depends on the properties of both the material of the top tip and the material of the inclined plane. For very low inclination angles the trajectory also depends on the radius of curvature of the tip of the top, while for larger inclination the behavior is independent of the shape of the top. The behavior undergoes a dynamic instability near $\tan\alpha=\mu_c$ and tops on strongly inclined planes accelerate downhill similar to other non spinning objects. We thus add another example of geometric physics to the rich collection of examples in this exciting field.  

\section{Acknowledgment}
We thank D. De las Heras for useful discussion.
We acknowledge funding by the Deutsche Forschungsgemeinschaft (DFG, German Research Foundation) under project number 440764520.

%
%
%
%
%


\begin{thebibliography}{0}
	\bibitem{Berry}
	M. V. Berry;  
	Quantal Phase Factors Accompanying Adiabatic Changes. \textit{Proc. Royal Soc. A.}, 1984,  \textbf{392}, 45–57. 
		
		\bibitem{Hannay}
	J. H. Hannay;
	Angle variable holomony in adiabatic excursion of an integrable Hamiltonian
	{\it J. Phys.}, 1985, {\bf A18},  221-230. 
	
		\bibitem{Haken}
	C.Z.Ning and H. Haken;
	Geometrical phase and amplitude accumulations in dissipative systems with cyclic attractors 
	\textit{Phys. Rev. Lett.}, 1992, \textbf{68}, 2109–2122.
	
	\bibitem{HighEPHys}
	K. Ranganathan, H. Sonoda and B. Zwiebach; 
	Connections on the state space over conformal field theories, 
	\textit{Nucl. Phys. B}, 1994, ßtextbf{414}, 405. 
	
	 \bibitem{SSP-TI} 
Yuanbo Zhang, Yan-Wen Tan, Horst L. Stormer, and Philip Kim; 
Experimental observation of the quantum Hall effect and Berry's phase in graphene
\textit{Nature}, 2005, \textbf{438}, 201–204. 


\bibitem{Hasan}
M. Z. Hasan, and C. L. Kane; 
Colloquium: Topological insulators.
{\it Rev. Mod. Phys.}, 2010,  {\bf 82}, 3045-3067. 

\bibitem{TI}
S-Q. Shen; 		
Topological insulators: Dirac equation in condensed matters,
{\it Springer Science, and Business Media} (2013).
	
		\bibitem{Foucault}
	Jens von Bergmann; HsingChi von Bergmann; Foucault pendulum through basic geometry. \textit{Am. J. Phys.}, 2007  \textbf{75} 888–892. 
	
		\bibitem{Wilczek1}
	A. Shapere, and F. Wilczek; 
	Geometry of self-propulsion at low Reynolds number. 
	{\it Journal of Fluid Mechanics}, 1989, {\bf 198}, 557-585.
	
	\bibitem{Wilczek2}
	A. Shapere, and F. Wilczek; 
	Efficiency of self-propulsion at low Reynolds number. 
	{\it Journal of Fluid Mechanics}, 1989, {\bf 198}, 587-599.

	\bibitem{fiber}   
    A. Tomita and R. Chiao;
  Observation of Berry's topological phase by use of an optical fiber
	{\it Phys. Rev. Lett.}, 1986 ,{\bf 57}, 937-940.
		
	\bibitem{Rechtsman}	M. C. Rechtsman,	J. M. Zeuner,	Y. 
	Plotnik,	Y. Lumer,	D. Podolsky,	F. Dreisow,	S. Nolte,	M. Segev, and A. Szameit;		
	Photonic Floquet topological insulators
	{\it Nature}, 2013, {\bfseries 496}, 196-200.	
	
	\bibitem{Mao} M. Xiao, G. Ma, Z. Yang, P. Sheng, Z.Q. Zhang, 
	and C.T. Chan;	
	Geometric phase and band inversion in periodic acoustic systems
	{\it Nature Phys.}, 2015, {\bfseries 11}, 240-244. 
	
	\bibitem{Murugan}
	A. Murugan and S. Vaikuntanathan;
	Topologically protected modes in non-equilibrium stochastic systems
	{\it Nat. Comm.}, 2017, {\bfseries 8}, 13881.
	
		\bibitem{Nash} L. M. Nash, D. Kleckner, A. Read, V. Vitelli, 
	Ari M. Turner, and W. T. M. Irvine;
	Topological mechanics of gyroscopic metamaterials
	{\it Proc. Nat. Acad. Sci.}, 2015, {\bfseries 112}, 14495-14500.
	
		\bibitem {Harland2018}	D. Harland, and N. S. Manton; 
	Rolling Skyrmions and the nuclear spin-orbit force
	{\it Nuclear Physics B}, 2018, {\bf 935}, 210-241.
	
	\bibitem{Rossi}
	A. M. E. B. Rossi, J. Bugase, and Th. M. Fischer;
	Macroscopic Floquet topological crystalline steel and superconductor  pump
	{\it Eur. Phys. Lett.}, 2017, {\bfseries 119}, 40001.
	
	\bibitem{tp1} J. Loehr, M. L\"onne, A. Ernst, D. de las Heras, and Th. M. Fischer; 
Topological protection of multiparticle dissipative transport.
{\it Nat. Comm.}, 2016, {\bfseries 7}, 11745.
	
	\bibitem{tp2}
	D. de las Heras, J. Loehr, M. L\"onne, and Th. M. \& Fischer;
	Topologically protected colloidal transport above a square magnetic lattice.
	{\it New J. Phys.} (2016), {\bfseries 18}, 105009.
	
	\bibitem{tp3}
	J. Loehr, D. de las Heras, M. L\"onne, J. Bugase, A. Jarosz, M. Urbaniak, F. Stobiecki, A. 
	Tomita,	R. Huhnstock, I. Koch, A. Ehresmann, D. Holzinger, and Th. M. Fischer;	 		
	Lattice symmetries and the topologically protected transport of colloidal particles.
	{\it Soft Matter} (2017), {\bfseries 13}, 5044-5075.
		
	\bibitem{Mahla}
	M. Mirzaee-Kakhki, A. Ernst, D.de las Heras, M. Urbaniak, F. Stobiecki, J.  G\"ordes, M. Reginka, A. Ehresmann, and Th. M. Fischer;
	Simultaneous polydirectional transport of colloidal bipeds
	{\it Nature Comm.}, 2020, {\bfseries 11}, 4670.
	
\bibitem{Popov}
Valentin l. Popov;
Kontaktmechanik und Reibung;
Ein Lehr- und Anwendungsbuch von der Nanotribologie bis zur numerischen Simulation
Springer Verlag Berlin Heidelberg (2009) 
	
\bibitem{STIP1}	
Richard J. Cohen;
The tippe top revisited
\textit{American Journal of Physics} 1977, \textbf{45}, 12.


\bibitem{STIP2}	
Ledo Stefanini;
Behavior of a real top
\textit{American Journal of Physics} 1979, \textbf{47}, 346.

\bibitem{STIP3}	
Rod Cross;
The rise and fall of spinning tops
\textit{American Journal of Physics} 2013, \textbf{81}, 280.

\bibitem{STIP4}	
Rod Cross;
Surprising Behavior of Spinning Tops and Eggs on an Inclined Plane
\textit{Phys. Teach.}, 2016, \textbf{54}, 28.

\bibitem{quietlife}
S. Hermann, D. de las Heras, and M. Schmidt;
Phase separation of active Brownian particles in two dimensions: anything for a quiet life
\textit{Mol. Phys.}, 2021, \textbf{}, e1902585,
DOI: 10.1080/00268976.2021.1902585 

\bibitem{turnover}
H. Ockendon, J. R. Ockendon, and T. Tokieda;
 The Turnover of a Tippy Top
\textit{J. Dyn. Diff. Equ.}, 2015, \textbf{27} 929-940.




	
\end{thebibliography}
\end{document}